\documentclass[twocolumn,prl,showpacs,preprintnumbers,amsmath,amssymb,superscriptaddress,floatfix]{revtex4}
\usepackage{graphicx}
\usepackage{xspace}
\usepackage{hyphenat}
\graphicspath{{../figures/}}

\begin{document}

\title{Light forces in ultracold photoassociation}
\author{E. Gomez, A. T. Black, L. D. Turner, E. Tiesinga, P. D. Lett}
\affiliation{National Institute of Standards and Technology,
Gaithersburg, Maryland 20899}

\date{\today}

\newcommand{\dwave}{$d$\hyp{}wave~}
\begin{abstract}
We study the time-resolved photoassociation of ultracold sodium in an
optical dipole trap. The photoassociation laser excites pairs of atoms
to molecular states of large total angular momentum at high
intensities (above 20\,kW/cm$^{2}$). Such transitions are generally
suppressed at ultracold temperatures by the centrifugal barriers for
high partial waves. Time-resolved ionization measurements reveal
that the atoms are accelerated by the dipole potential of the
photoassociation beam. We change the collision energy by varying the
potential depth, and observe a strong variation of the
photoassociation rate.  These results demonstrate the important role
of light forces in cw photoassociation at high intensities.
\end{abstract}


\maketitle

\section{Introduction}
Photoassociation spectroscopy is a flexible and precise tool for
determining molecular level structures~\cite{tiesinga05} and binding
energies~\cite{jones96}, as well as atomic lifetimes~\cite{mcalexander96}
and scattering lengths~\cite{gardner95}. In photoassociation (PA) a
photon excites a pair of colliding atoms into a bound molecular
level. The frequency of the excitation laser determines the
vibrational and rotational level of the molecule produced.
Typically, photoassociation spectroscopy of ultracold atoms is
performed at intensities on the order of the saturation intensity.
Studies of photoassociation at higher intensities have focused on PA
rate limits~\cite{prodan03, mckenzie02,kraft05}, rather than the
photoassociation spectrum.

We present a spectroscopic study of high-intensity photoassociation in
trapped, ultracold sodium atoms. The spectra reveal excitation to
molecular states of unexpectedly large angular momentum, given the
initial temperature of the atoms. These spectral features persist even
in a Bose-Einstein condensate, for which all but $s$-wave collisions
are ordinarily suppressed. At first glance, the excitation of these
high rotational states seems to be at odds with conservation of
angular momentum. We show that the excitation to states of large
angular momentum is explained by atoms accelerating in the attractive
dipole potential created by the photoassociation beam.

Photoassociative production of molecules into states carrying
nominally forbidden angular momenta has previously been observed and
explained by an array of causes. An experiment in a cesium
magneto-optical trap used the trapping laser to excite quasimolecules
at long range. There the resonant dipole-dipole interaction
provided sufficient kinetic energy to produce eight rotational
lines~\cite{fioretti99}. A similar experiment in sodium observed up to
22 lines arising from the same interaction~\cite{shaffer99}. In two-photon,
``ladder-type'' photoassociation in potassium, a forbidden excitation
to a doubly excited high-rotational state proceeded off-resonance
through a nearby rotational state~\cite{stwalley99}. As expected for
such off-resonant processes, the states of large angular momentum
appeared only at fairly high intensity ($\sim$~200 W/cm$^2$). In
contrast to the above results, the present work demonstrates the
production of surprisingly high angular momenta arising from a single
laser frequency at the photoassociation transition.

We obtain PA spectra showing the rotational series for a nearly
forbidden transition to a molecular level dissociating to the Na
$^2S + ^2P_{3/2}$ asymptote. At low photoassociation intensities,
the rotational spectra comprise transitions to molecular states of
low total angular momentum, as expected for the temperature of the
atomic sample. Increasing the PA intensity causes spectral features
corresponding to states of high angular momentum to appear.
Measuring the time-dependent PA rate with 1~$\mu$s resolution
reveals oscillations in the PA signal, as well as a prompt step in
the signal that depends on the rotational quantum number. We compare
the signal to a Monte Carlo model of atomic dynamics in the Gaussian
dipole potential of the PA beam. The comparison shows that
mechanical effects induced by the photoassociation laser produce the
high-angular-momentum peaks in the spectrum.

\section{Experimental setup}

We capture $^{23}$Na atoms in a crossed dipole trap as described
previously~\cite{dumke06}. We start with a dark-spot magneto-optical
trap (MOT) at 300~$\mu$K, loaded by a Zeeman slower, and transfer
the atoms into the dipole trap. A 100 W ytterbium fiber laser
focused to a $1/e^2$ waist of 60~$\mu$m captures the atoms in a
crossed beam configuration. After 10 seconds of MOT loading, we
switch on the dipole trap beams at full power for 50 ms and then
apply optical molasses for an additional 30 ms to improve the
loading into the dipole trap. We load 3~$\times$~10$^6$ atoms into
the crossed dipole trap at a temperature of 50~$\mu$K and a density
of 10$^{14}$ cm$^{-3}$. Ramping down the fiber laser power produces
a Bose-Einstein condensate (BEC) of 150\,000 atoms after 9 seconds.
At all stages the atoms are mostly unpolarized in the $F=1$
hyperfine state.

A photoassociation laser beam is focused through the atoms and produces an
ion signal. Most of the data presented here was obtained from thermal
atoms in the dipole trap, with some additional results obtained from
the MOT and BEC stages. The PA laser excites the atoms to the
rotational levels of the $v=48$ vibrational level of a 1$_g$
potential. At short internuclear separations, this 1$_g$ potential
correlates to the Na$_2$ $1^1\Pi_g$ Born-Oppenheimer potential, and at
large separations to the Na $^2$S + Na $^2$P$_{3/2}$ atomic limit.  A second
photon from the same laser ionizes the molecule produced (see inset to
Fig.~\ref{scanhighj}).  The experiment occurs in a metal vacuum
chamber and a microchannel plate operating at a potential of $-2$~kV
detects the ions produced, with a time resolution better than
1~$\mu$s. The PA transition to the 1$_g$ level is weakly allowed by
singlet-triplet mixing, but it still presents a relatively strong
one-color ionization signal. This level has been the subject of many
previous photoassociation
studies~\cite{lett93,ratliff94,napolitano94,jones99,jones97}.  The use
of a weak transition allows the resolution of rotational levels at
high intensities, which otherwise might be unresolved due to power
broadening and state mixing.

We tightly focus the single-mode, fiber-coupled PA laser to $1/e^2$
radii between 14 and 22~$\mu$m and use PA powers up to 370~mW,
giving peak intensities as high as $100$~kW/cm$^2$. For comparison,
previous high-intensity cw photoassociation studies have made use of
beam intensities up to
1~kW/cm$^2$~\cite{schloder02,mckenzie02,prodan03,kraft05}. The
Gaussian PA laser beam creates a dipole potential with a depth of
6~mK at the maximum intensity. As we describe later, we determine
the beam waist both from the motion of the atoms in the PA laser
potential, and directly using a beam profiler.

\section{Rotational spectrum}

\begin{figure}
\leavevmode \centering
\includegraphics[width=3.1in]{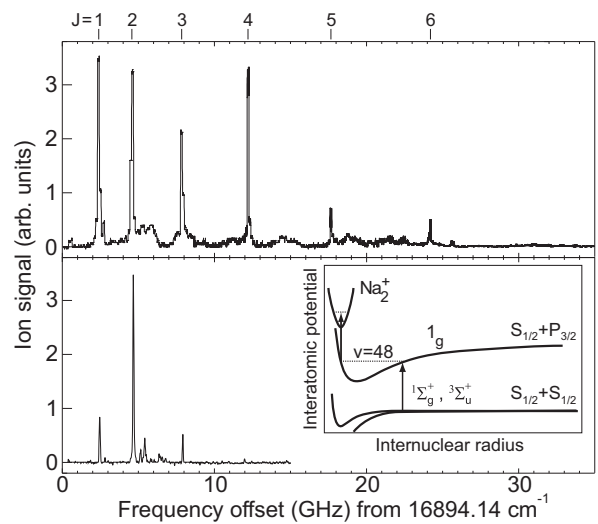}\caption{Rotational
spectrum of the $v$=48 vibrational level of the 1$_g$ potential in a
dipole trap. The lower (upper) trace shows the spectrum at low
(high) intensities. The initial temperature of the atoms is
approximately 50~$\mu$K, the upper trace PA laser power is 200~mW
and its waist is $22.3\pm 2.2$~$\mu$m. The markers above the peaks
show a fit to a rotational sequence running from $J=1$ to $J=6$. The
inset is a schematic energy level diagram showing the relevant
molecular potentials and indicating the photoassociation and
ionization processes. \label{scanhighj}}
\end{figure}

Figure~\ref{scanhighj} shows the ion signal spectra from
photoassociation in the crossed dipole trap.  We scan the frequency of
the PA laser and at each frequency apply a 10~ms pulse of PA light to
atoms initially at a temperature of 50~$\mu$K. The upper spectrum,
taken at a PA intensity of 26~kW/cm$^2$, shows a clear rotational
sequence up to $J=6$, with energy $E_\mathrm{rot}=B_v J(J+1)$, where
$B_v/h=540\pm40$~MHz is the rotational constant and $J$ is the
magnitude of the angular momentum. At these temperatures, only the
first three peaks are clearly visible in the lower spectrum, which was
taken at 1~kW/cm$^2$.

The incoming scattering state in the photoassociation process is
described by its collision energy $E$ and its partial wave $\ell$,
the magnitude of the orbital angular momentum of the atom pair
$\vec{\ell}$. For the excited $1_g$ molecules, the angular momentum
is given by $\vec{J}=\vec{L}+\vec{S}+\vec{\ell}$, where $\vec{L}$ is
the total electron orbital angular momentum and $\vec{S}$ is the
total electron spin. Electric-dipole transitions satisfy the
selection rule $\Delta\ell=0$; hence, the scattering state can only
be excited to rotational levels $J$ satisfying
\begin{equation}
\ell-|\vec{L}+\vec{S}| \leq  J \leq \ell+|\vec{L}+\vec{S}|.
\label{eq:ineq}
\end{equation}
We now consider the appropriate values for $\vec{L}$ and $\vec{S}$.

For individual atoms, the electronic orbital angular momentum is a
good quantum number. In molecules with well-separated electron
clouds, as in the current case, the total electronic orbital angular
momentum $L$ is given by the magnitude of the vector sum of the
atomic components. Thus $L=1$ for the excited $1_g$ molecule
dissociating to the S~+~P limit.

The $1_g$ potential in question is a mixture of singlet ($S=0$) and
triplet ($S=1$) character. In fact, the $v$=48 vibrational level of this
potential has about 99.5\% 1$^1\Pi_g$
character~\cite{movre77,napolitano94}. The remaining $\approx$0.5\% is
divided between the 1$^3\Sigma_g^{+}$ and 1$^3\Pi_g$ Born-Oppenheimer
states.

The electronic part of the scattering wavefunction is a linear
superposition of the X$^1\Sigma_g^+$ and a$^3\Sigma_u^+$ electronic
ground states. The precise linear superposition is determined by the
hyperfine state of the atoms. Selection rules for electric-dipole
transitions also require that only $u\leftrightarrow g$ and $\Delta
S=0$ transitions are allowed. Consequently, the $1_g$ vibrational
level can only be accessed through the a$^3\Sigma_u^+$ component of
the scattering state, giving $S=1$ in Eq.~\ref{eq:ineq}. The
triplet character of the $1_g$ state is small, making the transition
only weakly allowed. The natural line width of the transition
is 70~kHz~\cite{jones99}, and the observed width is limited by
hyperfine structure (unresolved in the current experiment) to
$\simeq$ 30~MHz~\cite{jones99}.

Inserting $L=1$ and $S=1$ into Eq.~\ref{eq:ineq} gives the partial
wave contributions for each rotational level $J$.  The lowest partial
wave contributions for $J$=1, 2, 3, 4, 5, and 6 are $s$, $s$, $p$,
$d$, $f$, and $g$, respectively. At low collision energies, the largest
contribution to the transition comes from the lowest allowed partial
wave.

The height of the centrifugal barrier for $p$ and $d$~waves in sodium
is 1.0~mK and 5.3~mK respectively~\cite{jones06}, much bigger than the
50~$\mu$K temperature of the atoms in the dipole trap. The outer
vibrational turning point of the $v$=48 state lies inside the
centrifugal barrier. Classically, this would preclude photoassociation
to this state for all but $s$~wave scattering.  Barrier tunnelling
predicts that $s$ and $p$~waves should contribute significantly to the
signal at the temperature of the atoms in the dipole trap (as shown in
the lower spectrum of Fig.~\ref{scanhighj}), but the upper spectrum in
Fig.~\ref{scanhighj} shows a clear signal even for $J=6$ ($g$~wave
collisions). In fact, unusually high rotational levels are observed
even for a Bose-Einstein condensate, in which only $s$-wave
contributions are expected. If we initiate forced evaporation in the
crossed dipole trap to produce a condensate~\cite{dumke06} and then
apply the PA laser, we see a signal up to $J=4$
($d$~waves). Acquisition of a full spectrum in the BEC was not
attempted due to the much lower repetition rate. The presence of high
partial waves depends strongly on the PA intensity; they are not
visible at low intensities.

The two-step, one-color ion signal described above depends in
principle on both the photoassociation and ionization steps (see
inset to Fig.~\ref{scanhighj}). However, the ionization step acts
only as a saturated probe of the excited state population. The
results presented here are not sensitive to modest power broadening
or structure in the ionization step.

The effect of the photoassociation and ionization step can be
separated using two-color photoassociation. In this case, the beams
from two independently tunable lasers are overlapped to form the PA
beam. We fix the frequency of one laser and scan the frequency of the
second laser across a rotational peak ($J=4$), with a scan range of
1.6~GHz. We repeat the scan for different frequencies of the first
laser, varying its frequency over a range of 600~MHz centered on the
$J=4$ transition.  The scan shows a signal only when either laser is
on resonance with the transition to the $J=4$ rotational level. The
absence of additional structure is consistent with the ionization step
acting only as a probe, and with the absence of significant
multiphoton processes leading to high angular momenta.

\section{Time dependence of the ion signal}

\begin{figure}
\leavevmode \centering
\includegraphics[width=3.1in]{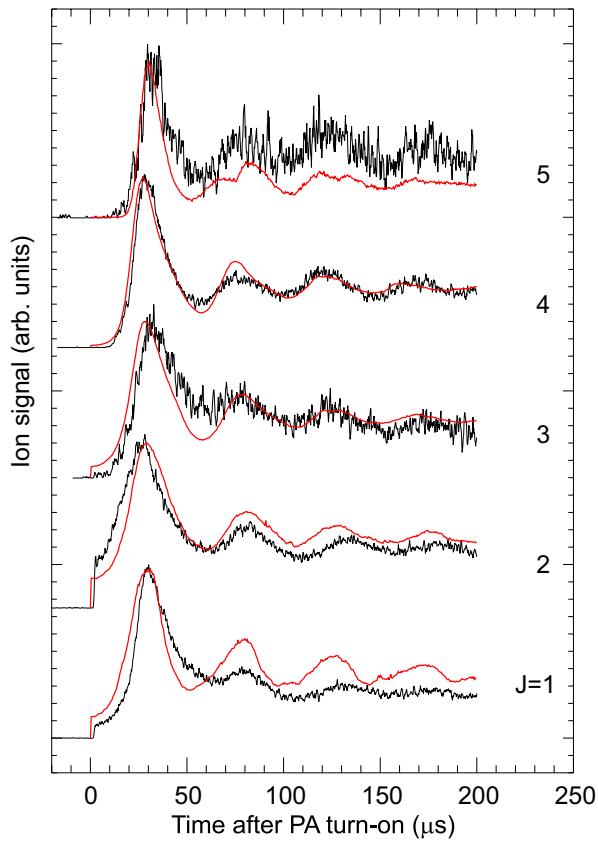}
\caption{(color online) Time dependence of the ion signal for different rotational
levels $J=1$ through $J=5$. The PA laser turns on at $t=0$. For
clarity the traces are displaced vertically.  All trace maxima have
been normalized to one. The data are shown in black, with simulated
signal overplotted in red. The initial temperature was 50\,$\mu$K, the dipole
trap $1/e^2$ radius was 85\,$\mu$m and the dipole trap depth was
500\,$\mu$K. The PA beam $1/e^2$ radius was 21\,$\mu$m. PA beam power
was nominally 250\,mW; the actual measured power for each peak was
used in the simulation. \label{timedependence}}
\end{figure}

To determine the mechanism responsible for the appearance of
high-$J$ features in the spectrum, we study the time dependence of
the ion signal for different rotational levels. We capture the atoms
in the crossed dipole trap and monitor the ion signal as a function
of time after we turn on the PA laser. Figure~\ref{timedependence}
shows the time dependence observed for different rotational levels,
each averaged over several trap loading cycles. The general
structure has an initial step, a prompt signal, which is present
only for low $J$ values, followed by several oscillations. At long
times (greater than 250~$\mu$s), the oscillations damp and the PA
rate becomes constant. The spectrum in Fig.~\ref{scanhighj} reflects
integration over this long-time rate. The time of flight of the ions
to the microchannel plate is below 1~$\mu$s, as confirmed by the
short delay between PA turn-on and the initial step in the ion
signal for the $J=1$ and $J=2$ traces. Each of the rotational lines
has between 10 and 16 hyperfine components, which are essentially
unresolved due to inhomogeneous energy broadening. The visibility of
the ion signal oscillations is higher when the PA laser is tuned to
the blue of the center of this hyperfine manifold, and lower when
tuned to the red. The ion signals shown in Fig.~\ref{timedependence}
were obtained with the PA laser tuned to produce the highest
integrated ion signal.

The oscillatory signal of Fig.~\ref{timedependence} suggests that the
photoassociation laser influences the motion of the atoms, which in
turn affects the photoassociation rate. As the PA laser is detuned red
of the sodium D lines, it forms an attractive dipole ``PA potential''.
Following the turn-on of the PA beam, this PA potential accelerates
atoms towards the beam center. Consequently, both the relative
collision energy and the atomic density increase rapidly at short
times and then oscillate. The oscillation period is close to $T/2$,
where $T$ is the period of a single atom in the harmonic part of the
potential. The period observed is consistent with a value of $T$
calculated from the atomic polarizability, and measured values of the
PA beam waist and power.  The PA rate increases with both collision
energy and density, and the two effects combine to give a maximum PA
rate after a time of roughly $T/4$.

The initial step in the ion signal (at $t=0$ in Fig.
\ref{timedependence}) is not related to the oscillations described, but
rather reflects the initial temperature and density of atoms in the
crossed dipole trap. The initial step is clearly visible for $J=1$ and
2 (due to $s$~wave collisions), is barely visible for $J=3$
($p$~waves), and is absent for $J=4$ ($d$~waves) and 5
($f$~waves). This is consistent with the initial ultracold
temperatures in the crossed dipole trap. We observe that the height of
the initial step of the $J=2$ peak is linear in the PA laser
intensity, indicating that the ionization transition is a saturated
probe of the PA rate.

The loss of visibility in the oscillations after a few periods is due
to dephasing in the anharmonic PA potential. The PA laser has a
Gaussian transverse profile and many atoms are far from the center
where the harmonic approximation holds. The signal also decays slowly
for long times as the atoms are lost from the dipole potential either
due to mechanical effects or photoassociation.

The observed oscillations are much faster than the characteristic time
for thermalization of the atom cloud~\cite{mudrich02}. In any case, no
oscillations are expected for a sample remaining in thermal
equilibrium as it evolves from one temperature to another.
Furthermore, increasing the temperature in a harmonic trap would reduce
the density of atoms; by contrast, we infer an initial increase in
density. Density oscillations due to the sudden addition of a deep
dimple potential were studied theoretically in Ref.~\cite{proukakis06}
in the context of atoms near condensation in a one-dimensional
gas. Despite having a thermal cloud, we infer similar density
oscillations from our photoassociation signal.

We also study the ion signals in the MOT in the absence of the
crossed dipole trap. The MOT is optimized for dipole trap loading,
with lower temperatures and lower densities than in our previous MOT
PA experiments~\cite{fatemi02}. Spectra taken in the MOT reveal a
rotational series similar to that measured in the crossed dipole
trap (Fig.~\ref{scanhighj}), with peaks up to $J=7$, despite the
sub-millikelvin MOT temperatures. Furthermore, no signal appears
when the PA beam illuminates the MOT for durations less than 1~ms.
This is the case even for the low rotational peaks $J=1$ and $J=2$.
However, for longer durations an ion signal does appear, reaching a
steady state after 30 ms. This cw signal is comparable in strength
to that in the crossed dipole trap, despite the initial density in
the MOT being of order $10^{10}$~cm$^{-3}$.

Apparently, atoms from the MOT are trapped in the PA potential over
time, reaching sufficient density to produce a visible PA signal only
after a few milliseconds delay. We test this interpretation by keeping
the PA laser on continuously, but detuned 40~MHz from the $J=2$ PA
resonance, to achieve steady-state trapping in the PA potential. As
expected, no ion signal is observed while the PA beam is
off-resonance. If the frequency of the PA beam is suddenly shifted
onto resonance, a steady ion signal appears instantly. No oscillations
similar to those in Fig.~\ref{timedependence} are observed.

If, in the above test, we suddenly double the PA beam power as we
shift the frequency onto resonance, the ion signal does exhibit
oscillations on a time scale similar to those in
Fig.~\ref{timedependence}. Atoms already trapped in the PA potential
suddenly gain kinetic energy, leading to the oscillating signal. This
shows that the mechanism responsible for high angular momentum peaks
in our MOT PA spectrum is the same as for the dipole trap experiments.

The advantage of working with the crossed dipole trap is that it
provides a high-density sample in which an ion signal appears
immediately after we turn on the PA beam. In this way we have good
control over the density and temperature of the sample before we
introduce the PA laser. Furthermore, the range of oscillation
frequencies of the atoms is smaller than in the MOT because the
crossed dipole trap is contained within one Rayleigh length of the
PA laser beam. Hence, the visibility of the oscillations in the
dipole trap is higher than in the MOT.

\section{Monte Carlo simulation of the ion signal}
We perform a classical Monte Carlo simulation of the ion signal as a
function of time. The key components of the simulation are a
description of the initial atomic positions and velocities; a
calculation of the motion of the atoms; and a model for the ion
signal, which makes use of the collisional energy dependence of the
PA rate.

We initialize the atoms with the position and velocity distributions
for the crossed dipole trap. We then introduce the PA potential and
calculate the resulting classical motion of the atoms. A typical
potential depth is around 2~mK, which is much larger than the
temperature of the atoms. The simulation exploits the cylindrical
symmetry of the experiment, and considers only the motion in the plane
perpendicular to the PA beam propagation direction. Motion along the
PA beam axis is neglected, as the dipole force is small in this
direction. We divide the radial coordinate into 100 bins and collect
all the atoms that fall into each bin as a function of time. We obtain
the relative velocity $v_{mn}$ between all possible pairs of atoms
$m$, $n$ in a particular bin to calculate the collision energy
$E_{mn}=\mu v_{mn}^2 / 2=\hbar^2 k_{mn}^2 / 2\mu$, with $\mu$ the
reduced mass. The simulated ion signal $S$ at a given time step in the
simulation is given by~\cite{napolitano94}
\begin{equation}
S(\omega-\omega_0)=\sum_i n_i N_i I_i \Big \langle \frac{\pi
v_{mn}}{k_{mn}^2}
\frac{Z(E_{mn})}{(\omega-\omega_0-\Delta)^2+\gamma_{mn}^2} \Big
\rangle_{mn}, \label{signaleq}
\end{equation}
where $n_i$, $N_i$, and $I_i$ are the density, atom number, and PA
laser intensity respectively for the bin $i$. The average runs over
all pairs of atoms in bin $i$. $Z(E)$ corresponds to the
intensity-normalized stimulated width for a collisional energy $E$ and
can be well approximated by the Wigner threshold law $Z(E)
\propto E^{\ell+1/2}$ at low energies. $\gamma_{mn}=\gamma_0+I_i Z(E_{mn})$ is the total width, with $\gamma_0$ the natural width.  $\omega$ is the frequency of the laser,
and $\Delta=AI_i-E_{mn}$ is the shift from the unperturbed resonant
frequency $\omega_0$ for the transition due to the collisional energy
$E_{mn}$ and the light shift $AI_i$. The signal $S$ is linear in $I$
at low intensities, due to the saturation of the ionization
transition.

The input parameters for the simulation are the initial
distributions of atoms in the dipole trap (determined by the dipole
trap waist, depth and the initial temperature), the power of the PA
laser beam, its beam waist and the light shift. We obtain the waist
from the measured PA laser power and the frequency of the
oscillating ion signal. The waist obtained this way is consistent
with the one measured directly with a scanning knife-edge beam
profiler. We measure the $s$-wave light shift with short PA pulse
durations on the $J=2$ peak, and use this experimentally-determined
light shift for all rotational peaks.

\section{Discussion}
The simulations confirm that the oscillations observed in the ion
signal are due to the motion of the atoms in the PA potential. This
motion causes a periodic modulation of the density at the bottom of
the potential, which is mainly responsible for the signal. There is
also an energy modulation, since the atoms gain a maximum kinetic
energy when they reach the bottom of the potential. The energy
increase is what makes it possible to see high rotational levels in
Fig. \ref{scanhighj}, which would otherwise be absent due to the
centrifugal barrier.

In an initial simulation, we assume Wigner threshold law behavior and
include only the lowest partial wave allowed for the rotational peak
in question.  This simple model reproduces the essential structure of
the oscillation, characterized by a strong initial peak followed by
further oscillations of diminishing amplitude. Quantitative agreement
is not, however, achieved and the height of the prompt ion signal is
not reproduced accurately.

To obtain quantitative agreement with the data we extended the
simulation to include the full hyperfine structure of each rotational
peak. In this extended model, the ion signal at each time step is
found by summing Eq.~\ref{signaleq} over the hyperfine substates $f$
of the rotational state in question,
\begin{equation}
S_\textrm{hfs}(\omega)=\sum_f S(\omega - \omega_f).
 \label{hfseq}
\end{equation}
Here $\omega_f$ is the unperturbed resonant frequency of each
hyperfine transition. A different function $Z_f(E)$ must be used for
the intensity-normalized stimulated width of each hyperfine
transition. This stimulated width is a sum over all contributing
partial waves and Zeeman sublevels of the colliding
atoms~\cite{tiesinga05,samuelis01}.

The simulated ion signal $S_\textrm{hfs}$ is shown in
Fig.~\ref{timedependence} superimposed on the experimental signal. The
amplitude of the simulated signal is normalized to the experiment. To
accurately reproduce the experimental procedure, the simulation was
performed for a range of laser tunings $\omega$, and the tuning
yielding the greatest integrated ion signal was selected. Because our
lasers have a nominal 5\,MHz stability, the simulated signal is a
weighted sum of the time series over an appropriate range of
detunings.

The simulation reproduces the oscillations, and also shows a prompt
ion signal due to $s$ and $p$ wave photoassociation. This prompt
signal is reproduced for $J=1$, 2 and 3, with the largest prompt
signal for $J=2$. The agreement between simulation and experiment is
best for $J=4$, possibly because the signal is dominated by a single
partial wave ($d$-wave). Lower rotational peaks have significant
contributions from two or more partial waves, and consequently are
more sensitive to the molecular structure model. The effect of the
dominant partial wave in each time series is manifest as the initial
peak becomes progressively steeper for angular momenta above $J=2$.

We also found that the simulation consistently showed reduced
visibility of the oscillations when the simulated laser frequency
was set red of the transition center, and enhanced visibility when
set to the blue.

In a further series of experiments, we explore different energy
regimes by repeating the measurement at various PA potential
depths. We increase the depth up to 6~mK by changing the laser
power. As we increase the depth of the potential, we observe an
increase in the oscillation frequency of the ion signal. The
oscillation frequency scales as the potential depth raised to the
power $0.55 \pm 0.09$, consistent with the square-root dependence
expected for oscillation in a Gaussian potential.

\begin{figure}
\leavevmode \centering
\includegraphics[width=3.1in]{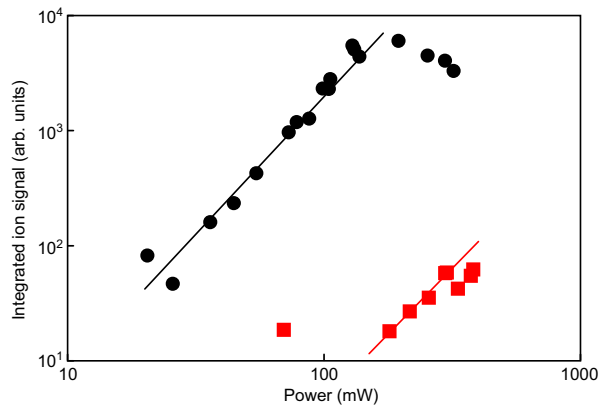}\caption{Power dependence of
the integrated ion signal for $J=4$ (black circles) and $J=5$ (red
squares). The lines correspond power law fits. The PA laser 1/e$^2$
radii is 14~$\mu$m. \label{powerboth}}
\end{figure}

We study the dependence of the magnitude of the ion signal on the
potential depth. Figure~\ref{powerboth} shows the integrated ion
signal as a function of the PA laser power for the $J=4$ (circles) and
$J=5$ (squares) peaks. The $J=5$ signal is considerably weaker and for
low powers it is buried in the noise. The signal follows approximately
a power law.  At high powers we observe a saturation of the signal
that happens at a lower power for $J=4$ than for $J=5$.

The PA lineshape suffers considerable modifications at high
powers. The observed shifts become larger than the line width at low
powers, and the spectrum is power broadened.  This suggests that the
saturation of the signal in Fig.
\ref{powerboth} is a consequence of the line shape modifications,
which affect the denominator of Eq.~\ref{signaleq}.

Simulations support the hypothesis that this saturation is due to
power broadening of the PA transition, with simulated ion signals
saturating as power broadening becomes significant. Further, in
simulations this saturation occurs at higher intensities for $J=5$
than for $J=4$, as expected from the behaviour of the stimulated width
function $Z_f(E)$. The simulation produces a power-law dependence, but
with a different exponent to that observed experimentally. The
accuracy of the molecular structure input is much more important for
the power dependence than it is for the time dependence of the signal.

\section{Conclusion}
In summary, we study the mechanical effects present in high
intensity photoassociation. The time resolution of the ion signal
provides clear evidence for oscillations of the atoms inside the PA
laser potential. We present a simulation that closely reproduces the
experimental results. The PA potential gives the atoms enough energy
to excite higher rotational levels, an excitation which is strongly
forbidden by centrifugal barriers in an unperturbed thermal gas at
the dipole trap temperature. In particular, when we start with a
Bose-Einstein condensate, we observe partial wave contributions up
to $d$ waves. The potential created by the PA laser imparts
collision energy to the atoms in a controlled, albeit
non-monochromatic way. The present results additionally demonstrate
the use of time-resolved photoassociation to study rapid density
dynamics in an atomic cloud~\cite{proukakis06}. The understanding of
mechanical effects will play an important role in future studies of
ultracold photoassociation at high intensities.

We thank Fredrik Fatemi for providing his earlier observations of
high rotational levels at high intensity.


\end{document}